\documentclass[reprint,amsmath,smsymb,aps,superscriptaddress,aip,apl,nofootinbib]{revtex4-2}
\usepackage{geometry}
\usepackage{titlesec}
\usepackage{bm}
\usepackage{gensymb}
 \usepackage{graphicx}
 \usepackage{multirow}
 \usepackage{tensor}
 \usepackage{verbatim}
 \usepackage{courier}
 \usepackage{mathtools}
  \usepackage[inline]{enumitem}
\graphicspath{ { } }
\usepackage{color}
\usepackage{amsmath}
\usepackage{lineno}
\usepackage{xfrac}
\usepackage[table]{xcolor}
\usepackage{appendix}
\usepackage{hyperref}

\begin{document}
\title{Evidence for subdominant multipole moments and precession in merging black-hole-binaries from GWTC-2.1}
\author{Charlie Hoy}
\affiliation{School of Physics and Astronomy, Cardiff University, Cardiff, CF24 3AA, United Kingdom}
\author{Cameron Mills}
\affiliation{School of Physics and Astronomy, Cardiff University, Cardiff, CF24 3AA, United Kingdom}
\affiliation{Albert-Einstein-Institut, Max-Planck-Institut for Gravitationsphysik, D-30167 Hannover, Germany}
\affiliation{Leibniz Universitat Hannover, D-30167 Hannover, Germany}
\author{Stephen Fairhurst}
\affiliation{School of Physics and Astronomy, Cardiff University, Cardiff, CF24 3AA, United Kingdom}
\date{\today}

\begin{abstract}
The LIGO--Virgo--KAGRA collaborations (LVK) produced a catalogue containing gravitational-wave (GW) observations from the first half of the third GW observing run (O3a). This catalogue, GWTC-2.1, includes for the first time a number of \emph{exceptional} GW candidates produced from merging black-hole-binaries with unequivocally unequal component masses. Since subdominant multipole moments and spin-induced orbital precession are more likely to leave measurable imprints on the emitted GW from unequal component mass binaries, these general relativistic phenomena may now be measurable. Indeed, both GW190412 and GW190814 have already shown conclusive evidence for subdominant multipole moments. This provides valuable insights into the dynamics of the binary. We calculate the evidence for subdominant multipole moments and spin-induced orbital precession for all merging black-hole-binaries in GWTC-2.1 that were observed during O3a and show that (a) no gravitational-wave candidate has measurable higher order multipole content beyond $\ell = 3$, (b) in addition to the confident subdominant multipole measurements in GW190412 and GW190814, GW190519\_153544 and GW190929\_012149 show marginal evidence for the $(\ell, |m|) = (3, 3)$ subdominant multipole, (c) GW190521 may have measurable subdominant multipole content and (d) GW190412 may show evidence for spin-induced orbital precession.

\end{abstract}

\maketitle

\section{Introduction}

Between 2015 and 2017, the Advanced LIGO~\cite{TheLIGOScientific:2014jea} (aLIGO) and Advanced
Virgo~\cite{acernese2014advanced} (AdV) gravitational-wave (GW) detectors performed their
first and second GW observing runs (O1 and O2). During this time, the LIGO Scientific and Virgo collaborations (LVC) announced GWs originating from ten binary-black-hole (BBH) mergers~\cite{Abbott:2016blz, TheLIGOScientific:2016pea, LIGOScientific:2016sjg, LIGOScientific:2017bnn, LIGOScientific:2017vox, abbott2017gw170814, abbott2019gwtc} and a single binary neutron star~\cite{abbott2017gw170817}. Additional compact binary candidates have also been reported by other groups~\citep{nitz20202, venumadhav2020new, zackay2019highly, zackay2019detecting}. 

The observed signals match well with the predictions of general relativity, calculated using accurate waveforms modelled to fit predictions from post-Newtonian methods and numerical relativity simulations.  However, two important general relativistic effects, higher order multipoles~\cite{Thorne:1980ru}
and spin-induced orbital precession~\cite{Apostolatos:1994mx} were not clearly identified in the signals observed during O1 and O2~\cite{Payne:2019wmy, abbott2019gwtc, fairhurst2019will}.%
\footnote{See Refs.~\cite{Chatziioannou:2019dsz,Mateu-Lucena:2021siq} for a discussion regarding marginal evidence for higher order multipoles in GW170729~\cite{abbott2019gwtc}, and Ref.~\cite{Chia:2021mxq} for discussion of evidence for higher modes and precession in GW151226.}
The gravitational wave emitted during a binary merger can be decomposed into a set of spin-weighted spherical harmonics~\cite{Thorne:1980ru}.  The majority of the gravitational wave signal is contained in the quadrupole mode, $(\ell, m) = (2, 2)$.  
Higher order multipoles are present for all binary systems, but they are typically at a much lower amplitude than the quadrupole~\cite[see e.g.][]{Mills:2020thr}. Spin-induced orbital precession arises when there is a misalignment between
the orbital angular momentum and the spins of each compact object leading to characteristic modulations in the amplitude and phase of the observed GW signal~\cite{Apostolatos:1994mx}. Including both higher order multipoles and precession in waveform models that are used to infer source properties through Bayesian inference~\cite[see e.g.][]{Veitch:2014wba, Lange:2018pyp, Ashton:2018jfp} can improve parameter measurement accuracy and provide additional constraints on the in-plane spin components of the binary~\cite[see e.g.][]{London:2017bcn, Kalaghatgi:2019log,LIGOScientific:2020stg, Abbott:2020khf,Green:2020ptm}. The importance of both
of these effects increase as the binary's mass ratio ($q = m_{1} / m_{2} \geq 1$)
increases~\cite{Mills:2020thr, Pekowsky:2012sr, Varma:2014jxa, Bustillo:2015qty, Capano:2013raa, vecchio2004lisa, Lang:1900bz, Pratten:2020igi, Green:2020ptm}. Clear evidence for asymmetric masses was absent in the binaries detected during O1 and O2~\cite{abbott2019gwtc}, making the observation of either precession and higher order multipoles challenging. 

An initial analysis of the first 6 months of data from the first half of the third GW observing run (O3a) by the LIGO-Scientific, Virgo and KAGRA collaborations (LVK) revealed a further 39 GW candidates described in the second GW catalogue (GWTC-2)~\cite{abbott2020gwtc}. Subsequent searches over the same period revealed a number of subthreshold candidates~\cite{Nitz:2021uxj, LIGOScientific:2021usb}, with the extended second GW catalogue, (GWTC-2.1), increasing the number of high-significance GW candidates observed during O3a to a total of 44~\cite{LIGOScientific:2021usb}. Recently, the LVK released the third GW catalogue (GWTC-3), which analysed the second half of the third GW observing run (O3b) and found a further 35 GW candidates~\cite{LIGOScientific:2021djp}.

In contrast to O1 and O2, several events in O3 had unequivocally unequal masses. First among these is GW190412~\cite{LIGOScientific:2020stg}, with a mass ratio of $\sim$4:1. The unequal mass ratio resulted in more significant higher order multipoles, and for the first time, imprints of subdominant multipole radiation oscillating at three times the orbital frequency -- the $(\ell, m) = (3, 3)$ multipole%
\footnote{($\ell$,m) should everywhere be read as ($\ell,|m|$) unless otherwise indicated.}
 -- were visible. Similarly, it was the first time that the in-plane spins, which lead to precession in the system, were constrained away from the prior~\cite{LIGOScientific:2020stg,Colleoni:2020tgc,Islam:2020reh,Nitz:2021uxj,Capano:2020dix}. Several months later GW190814 was detected with highly asymmetric component masses ($\sim$9:1) and a secondary component with a mass larger than any previously discovered neutron star and lighter than any black hole~\cite{Abbott:2020khf}. GW190814 had significant evidence of the (3,3) multipole~\cite{Abbott:2020khf, Capano:2020dix} and the most restrictive measurement of the in-plane spin components of any event observed to date. A combination of the higher order multipoles and the lack of evidence for precession reduced the uncertainty on the mass of the smaller object.

By comparing parameter estimates that were obtained with waveform models that a) included higher order multipoles and b) excluded higher order multipoles beyond $\ell = 2$, it has been suggested that several GW signals in O3a show possible evidence for higher order multipoles~\cite{abbott2020gwtc}. Similarly, by comparing the posterior and prior distributions for parameters characterizing spin-induced orbital precession, it has previously been shown that no event in O3a or O3b unambiguously exhibits spin-induced orbital precession~\cite{abbott2020gwtc, LIGOScientific:2021djp} (although see Ref.~\cite{Hannam:2021pit} which discusses strong evidence for precession in a GW candidate observed in O3b), but there is evidence that the observed population of BBHs have spins which are misaligned with the orbital angular momentum~\cite{abbott2020gwtc2prop, LIGOScientific:2021psn}. Other studies have investigated the evidence for higher order multipoles and precession for GW190412 and GW190814~\cite{LIGOScientific:2020stg, Abbott:2020khf,Colleoni:2020tgc,Islam:2020reh,Capano:2020dix}.  To date,  there has not been a study which quantifies the evidence for precession and higher order multipoles across all events on an event by event basis.

In this paper, we take advantage of the multipole decomposition for identifying the presence of higher order multipoles~\cite{Mills:2020thr} and the two-harmonic formalism for identifying the presence of precession~\cite{Fairhurst:2019_2harm, fairhurst2019will, Green:2020ptm} to establish if any BBH candidates in GWTC-2.1, that were observed during O3a, exhibit evidence for higher order multipoles and precession. We calculate the signal-to-noise ratio (SNR) in the $(\ell, m) \in \{(2,1), (3, 3), (4, 4)\}$ multipoles and from precession for the latest BBH observations and compare then to the expected distribution from noise. We include GW190814 since it is most likely (71\%) the result of a BBH merger~\cite{Abbott:2020khf}. We show that,

\begin{enumerate}
    \itemsep0em
    \item There is minimal evidence for GW emission in the subdominant multipole moments beyond $\ell = 3$,
    \item GW190814 has the largest evidence for the $(\ell, m) = (3, 3)$ multipole for all events in O3a with SNR in the $(3, 3)$ multipole $\rho_{33} = 6.2^{+1.3}_{-1.5}$,
    \item GW190814 and GW190412 show significant evidence for the $(3, 3)$ subdominant multipole moment while the evidence for GW190519\_153544 and GW190929\_012149 is marginal,
    \item GW190521 may show evidence for subdominant multipole moments. The reanalysis of GW190521 by Nitz \emph{et al.}~\cite{Nitz:2021uxj} suggests evidence for the $(3, 3)$ multipole while the initial analysis by the LVK shows minimal evidence,
    \item GW190412 and GW190915\_235702 show marginal evidence for spin-induced orbital precession while the population shows no significant evidence of precession.
\end{enumerate}

This paper is structured as follows: Section~\ref{sec:method} details the method used to infer the presence of higher order multipoles and precession in the observed GW data. We provide a brief overview of the two-harmonic approximation and a summary of how the SNR from precession and each subdominant multipole is calculated. We then explain how we construct the expected noise distribution for both measurements. In Section~\ref{sec:results} we present our results and indicate which GW events show evidence for subdominant multipoles and precession. We then conclude, in Section~\ref{sec:discussion}, with a summary and discussion of future directions.

\section{Method}
\label{sec:method}

\subsection{Calculating the SNR in precession and higher multipoles}
\label{sec:calc_rho}

In general relativity, GWs are fully described by two polarizations.
These polarizations can be decomposed into multipole moments using the $-2$ spin-weighted spherical harmonic orthonormal basis~\cite{goldberg1967spin}.
Coalescing compact binaries (CBCs) predominantly emit radiation at twice the orbital frequency in the leading order (2,2) multipole. The most important subdominant multipole for most CBCs is the (3,3) multipole, though the (4,4) multipole can be more significant for binaries whose components have comparable masses \cite{Mills:2020thr}.

Previous studies have identified evidence for subdominant multipole moments by either a) calculating Bayes factors, the difference in Bayesian evidences, through multiple parameter estimation studies~\cite[e.g.][]{Chatziioannou:2019dsz, Colleoni:2020tgc, Islam:2020reh} or via likelihood re-weighting~\cite{Payne:2019wmy}, b) statistically comparing posteriors obtained with waveform models which included higher order multipoles and those which excluded higher order multipoles beyond $\ell = 2$~\cite{Abbott:2020niy}, c) analysing time–frequency tracks in the GW strain data~\cite{LIGOScientific:2020stg,Roy:2019phx}, d) identifying if there is a loss in the observed coherent signal energy when comparing the output from the cWB detection pipeline~\cite{Klimenko:2015ypf} with predictions from a waveform model which excludes subdominant multipole moments~\cite{Abbott:2020khf}, or e) directly calculating the SNR of each ($\ell, m$) multipole $\rho_{\ell m}$~\cite{Mills:2020thr, LIGOScientific:2020stg, Abbott:2020khf, LIGOScientific:2021qlt}. Here, we identify whether multipoles other than the dominant (2,2) multipole have been observed by calculating the orthogonal SNR of each ($\ell, m$) multipole. This is achieved by decomposing a GW signal into each subdominant multipole, extracting the component that is orthogonal to the (2, 2) quadrupole and calculating the SNR associated with the resulting waveform.  

A binary with spin angular momenta $\mathbf{S}_{1}$ and $\mathbf{S}_{2}$ undergoes spin-induced orbital precession when the total spin $\mathbf{S}=\mathbf{S}_{1} + \mathbf{S}_{2}$ of the binary is misaligned with the Newtonian orbital angular momentum $\mathbf{L}$. In most cases, precession of the orbital plane leads to $\mathbf{L}$ precessing around the
approximately constant $\mathbf{J} = \mathbf{L} + \mathbf{S}$, leading to characteristic modulations in
the emitted GW signal~\cite{Apostolatos:1994mx,Kidder:1995zr}.

Previous studies have identified evidence for spin-induced orbital precession by either a) calculating Bayes factors~\cite[e.g.][]{LIGOScientific:2020stg, Abbott:2020khf,Abbott:2020mjq,Colleoni:2020tgc}, b) statistically comparing posterior distributions for parameters characterizing precession to their prior distributions~\cite[e.g.][]{abbott2019gwtc, Abbott:2020niy} or c) directly calculating the precession SNR, described as the contribution to the total SNR of the system that can be attributed to precession~\cite{Fairhurst:2019_2harm,fairhurst2019will,Green:2020ptm, Pratten:2020igi, LIGOScientific:2020stg, Abbott:2020khf,Abbott:2020niy}. In this paper, we identify if spin-induced orbital precession has been observed from a GW signal by calculating the precession SNR. The precession SNR $\rho_{\mathrm{p}}$ is calculated by decomposing the (2,2) multipole into non-precessing harmonics, whose frequencies differ by multiples of the precession frequency, and isolating the SNR contained in the second most significant harmonic (see \cite{Fairhurst:2019_2harm} for details). 
If $\rho_{\mathrm{p}}$ is small,
the amplitude of the second harmonic is
insignificant the observed waveform carries little imprint of orbital precession. In this case, we would
observe a GW signal which looks like the dominant non-precessing harmonic. The precession SNR has been shown previously to accurately identify whether spin-induced orbital precession has been observed in simulated GW signals~\cite{Green:2020ptm,Pratten:2020igi}.

The formalisms for $\rho_{\ell m}$ and $\rho_{\mathrm{p}}$ were initially developed for waveform models containing higher order multipoles~\cite{Mills:2020thr} or precession \cite{Fairhurst:2019_2harm,fairhurst2019will} respectively, but not both.  In this paper, we apply them to posterior samples obtained with a gravitational waveform model containing both higher multipoles and precession. We heuristically justify this by noting that, in most cases, both precession and higher multipoles are small corrections to the leading order gravitational waveform.  It is therefore reasonable to expect that the precession correction to the higher multipoles will be an even smaller effect which can be safely ignored. We leave a detailed derivation of this for future work~\cite{Fairhurst:2019_2harm}. Briefly, a GW signal containing only the $(\ell,m)=(2,2)$ multipole can be written as a sum of 5 non-precessing harmonics, $h_{22} = \mathcal{A}_{22,0}h^{22,0} + \mathcal{A}_{22,1}h^{22,1} + \ldots$,  where $\mathcal{A}_{22,n}$ is the (complex) amplitude of the $n$th harmonic $h^{22,n}$ which depends upon the orientation of the signal.  The amplitude of each harmonic scales with $b^{n}$ where $b = \tan{\beta/2}$ and $\beta$ is the opening angle (the polar angle between $\mathbf{L}$ and $\mathbf{J}$). The characteristic amplitude and phase modulations associated with precession are caused by the beating of these 5 non-precessing harmonics. Since for the majority of cases $b \ll 1$, a GW signal containing only the $\ell=2$ multipoles can approximately be written as a sum of the two leading harmonics, $h_{22}\approx \mathcal{A}_{22,0}h^{22,0} + \mathcal{A}_{22,1}h^{22,1}$. When the GW signal includes other multipoles, they can be decomposed similarly.  For example, we can express $h_{33} \approx \mathcal{A}_{33,0}h^{33,0} + \mathcal{A}_{33,1}h^{33,1}$
where $\mathcal{A}_{33,n}$ is the amplitude of the $n$th harmonic $h^{33,n}$. As before, the amplitude of the precession harmonics scale as $b^{n}$.  Furthermore, the overall amplitude of the (3,3) multipole is typically much lower than the (2,2) multipole.  Therefore, to a good approximation, we can write the waveform as
$h\approx \mathcal{A}_{22,0}h^{22,0} + \mathcal{A}_{22,1}h^{22,1} +  \mathcal{A}_{33,0}h^{33,0}$.
The $(\ell,m) = (4,4)$ precession multipoles can be added in a similar fashion, although their amplitude is generally smaller than the (3,3) and can often be neglected.  When computing $\rho_{\mathrm{p}}$ throughout this paper, we only consider the precession power in the leading $(2,2)$ multipole. Similarly, when calculating the power in higher multipoles, we consider only the contribution from, e.g., $h^{33,0}$ and neglect the precession corrections.  

\subsection{Calculating the expected distribution of $\rho_{p}$ and $\rho_{\ell m}$ in the absence of a signal}
\label{sec:noise_rho}

In order to assess the significance of precession and higher order multipoles, we compare the inferred $\rho_{\mathrm{p}}$ and $\rho_{lm}$ distributions to the expected distribution from noise. Since the statistical properties of $\rho_{\mathrm{p}}$ and $\rho_{lm}$ are similar, the expected noise distribution has the same functional form for both measurements. Below we summarize the derivation of the common noise distribution (parameterized by $\rho$ which denotes either $\rho_{\mathrm{p}}$ or $\rho_{lm}$) and we refer the reader to Refs.~\cite{Green:2020ptm, Mills:2020thr} for further details.

We consider a gravitational waveform comprising a dominant contribution $h_{0}$, the leading precession harmonic of the (2,2) multipole, and a single, additional, subdominant contribution $h_{1}$ arising either from a higher multipole or from precession.  The gravitational waveform can be written as
\begin{equation}
h = \mathcal{A}_{0}(\bm{\lambda})h_{0}(\bm{\lambda}) + \mathcal{A}_{1}(\bm{\lambda})h_{1}(\bm{\lambda})
\end{equation}
where $\mathcal{A}_{i}(\bm{\lambda})$ are overall amplitudes which depend upon the orientation of the binary, and $\bm{\lambda}$ denotes the parameters of the system~\cite{Fairhurst:2019_2harm, Mills:2020thr}. The GW likelihood may then be factorized into two components: one describing the contribution from the dominant harmonic, $\Lambda_{0}(\bm{\lambda})$, and another describing the contribution from the subdominant harmonic, $\Lambda_{1}(\bm{\lambda})$.
\begin{widetext}
    \begin{eqnarray}
        p(d|\bm{\lambda})
        &\propto& \exp\left(-\frac{1}{2}
        \langle d - \left[\mathcal{A}_{0}(\bm{\lambda})h_{0}(\bm{\lambda}) + \mathcal{A}_{1}(\bm{\lambda}){h}_{1}(\bm{\lambda}) \right] |
        d - \left[\mathcal{A}_{0}(\bm{\lambda})h_{0}(\bm{\lambda}) + \mathcal{A}_{1}(\bm{\lambda}){h}_{1}(\bm{\lambda}) \right] \rangle \right) \\
        &\propto& \exp\left( \mathcal{A}_{0}(\bm{\lambda})
          \langle d | h_{0}(\bm{\lambda}) \rangle
         - \frac{|\mathcal{A}_{0}(\bm{\lambda})|^{2}}{2} \langle h_{0}(\bm{\lambda}) | h_{0}(\bm{\lambda}) \rangle\right)
         \times \exp\left(
         \mathcal{A}_{1}(\bm{\lambda})
         \langle d |  h_{1}(\bm{\lambda}) \rangle
         - \frac{|\mathcal{A}_{1}(\bm{\lambda})|^{2}}{2} \langle h_{1}(\bm{\lambda}) | h_{1}(\bm{\lambda}) \rangle
         \right) \nonumber \\
        &=:&\Lambda_{0}(\bm{\lambda}) \times \Lambda_{1}(\bm{\lambda})\, \nonumber
    \end{eqnarray}
\end{widetext}
where $d$ is the GW strain data. In the second line, we have absorbed the (constant) $\langle d | d \rangle$ term into the proportionality, and we have assumed that the dominant and subdominant harmonics are orthogonal $\langle h_{0} | h_{1} \rangle = 0$.%
\footnote{The calculation can be fairly simply generalized to the case where the harmonics are not orthogonal by simply replacing $h_{1}$ by $h_{1}^{\perp}$, the component of $h_{1}$ orthogonal to $h_{0}$ \cite{Mills:2020thr}. For ease of presentation, all calculations are in terms of $h_{1}$ but when we calculate $\rho_{\ell,m}$ and $\rho_{\mathrm{p}}$ in this work, we use $h_{1}^{\perp}$.}

The phase evolution of the gravitational waveform, encoded in $h_{0,1}(\boldsymbol{\lambda})$, is determined by the masses and aligned spin components of the binary.  Since the dominant harmonic has the largest SNR, its measurement will primarily be used to determine the evolution of the waveform.  An observation of the sub-dominant harmonic will provide a small improvement to the evolution of the waveform. However, for simplicity, we neglect it in the following discussion.  In this case, the subdominant harmonic $h_{1}$ is known and only the overall amplitude and phase, encoded in $\mathcal{A}_{1}$ remain to be determined.  Furthermore, the value of $\mathcal{A}_{1}$ will typically be unconstrained by the observation of $\mathcal{A}_{0}$ --- in the case of precession, both the amplitude and phase of $\mathcal{A}_{1}$ are free as they depend upon the in-plane spins, while for higher multipoles the amplitude and phase of $\mathcal{A}_{1}$ will depend upon the orientation and mass ratio of the binary which are generally not precisely measured.
Therefore, in the simplest approximation, we can simply maximize the likelihood $\Lambda_{1}(\boldsymbol{\lambda})$ over $\mathcal{A}_1$,
\begin{equation}
    \Lambda_{1}(\boldsymbol{\lambda})_{\mathrm{max}} = \exp\left(\frac{[\rho_1^{\mathrm{MF}}]^{2}}{2}\right)
\end{equation}
where the matched filter SNR, $\rho_1^{\mathrm{MF}}$ is defined as
\begin{equation}
[\rho_1^{\mathrm{MF}}]^{2} = 
\frac{\left[
(s | h_{1})^2  +
(s | i h_{1})^2
\right]}{|{h}_{1}|^{2}} \, .
\end{equation}
$[\rho_1^{\mathrm{MF}}]^{2}$ will be chi-squared distributed with two degrees of freedom in the absence of signal, and non-centrally chi-squared distributed in the presence of a signal.  Using the maximum likelihood, a threshold of $\rho_{1} \ge 2.1$ is a requirement for observation, at 90\% confidence, of precession or higher multipoles (see Refs.~\cite{Fairhurst:2019_2harm, Mills:2020thr} for details).

Alternatively, we can marginalize the likelihood, $\Lambda_{1}(\boldsymbol{\lambda})$, over the unknown phase to obtain a likelihood as a function of $\rho_{1}$ as
\begin{equation}
\label{eq:marg_like}
    \Lambda_{1}(\rho_{1}) \propto  I_{0}(\rho_1^{\mathrm{MF}}\, \rho_{1}) \exp\left( - \frac{[\rho_1^{\mathrm{MF}}]^{2} + \rho_{1}^{2}}{2} \right) \, .
\end{equation}
Here, $I_{0}$ is the Bessel function of the first kind. The expected posterior distribution for $\rho_{1}$ is, therefore,
\begin{equation}\label{eq:analytic_posterior}
    p(\rho_{1} | d) \propto p(\rho_{1}) \, \Lambda_{1}(\rho_{1}) \, ,
\end{equation}
where $p(\rho_{1})$ is the prior distribution for $\rho_{1}$.  
For the case of uniform priors on the complex amplitude $\mathcal{A}_{1}$, $p(\rho_{1} | d)$ takes the form of a non-central $\chi$ distribution with 2 degrees of freedom with non-centrality parameter equal to $\rho_1^{\mathrm{MF}}$.

As shown in Refs.~\cite{Mills:2020thr, Green:2020ptm}, we can obtain a better prediction for the posterior distribution $p(\rho_{1} | d)$, by using the measurement of the dominant harmonic, $h_{0}$, to place a more informative prior on the strength of the sub-dominant harmonic. For instance, an observation of a close to equal mass or face-on binary reduces the expectation of observing higher harmonics or precession. Here, we construct an \emph{informed prior} for $p(\rho_{1})$ using results, where available, from a parameter estimation analysis that includes only the dominant multipole. This \emph{informed prior} is what results from calculating $p(\rho_{1} | d)$ in Eq.~(\ref{eq:analytic_posterior}) while taking $\Lambda_{1}(\rho_{1})=1$.%
\footnote{For precession there are additional parameters that must be marginalized over which are not inferred with dominant multipole models: the precession spin $\chi_p$~\cite{Schmidt:2014iyl} and the precession phase. However, the inference of aligned spin and mass ratio does provide additional constraints on these parameters, and so rather than assuming the default prior on these parameters, we condition on the measured aligned spin and mass ratio.}

\subsection{Assessing the significance of precession and higher multipoles}
\label{sec:signif_rho}

We follow the method introduced in Refs.~\cite{Green:2020ptm,Mills:2020thr} and calculate $\rho_{\ell m}$ and $\rho_{\mathrm{p}}$ from the inferred properties of each compact binary merger in O3a. To do this we use posterior samples from the GWTC-2.1 data release~\cite{ligo_scientific_collaboration_and_virgo_2021_5117703}. For GW candidates not included in the GWTC-2.1 data release, samples from the GWTC-2 data release~\cite{ligo_scientific_collaboration_and_virgo_2021_gwtc2} are used. Further details about the specific posterior samples used are in Appendix \ref{ap:samples}. We use the parameters of each sample to generate the leading order precession and higher multipole contributions to the waveform and calculate $\rho_{\ell m}$ and $\rho_{\mathrm{p}}$ for the network at the time of the event. This calculation uses the conversion module publicly available in {\sc{PESummary}}~\cite{Hoy:2020vys}.%
\footnote{When computing $\rho_{\ell m}$ we generate a non-precessing waveform (i.e. set the in-plane spin components to zero) and calculate the higher multipole SNR for that waveform.  This is not exactly identical to the prescription above but, provided that the $b \ll 1$, any differences will be small.  In particular, precession adds a secular drift in the phase evolution of the waveform.  However, as we only use the waveform to calculate the expected SNR in the higher multipoles (i.e. we do \textit{not} matched filter against the data), this small phase difference will not impact the value of $\rho_{33}$.}

To assess the significance of precession and higher multipoles in gravitational-wave signals, we calculate the probability that the observed $\rho_{\ell m}$ and $\rho_{\mathrm{p}}$ are caused by noise. In other words, we calculate the probability that the inferred $\rho_{\ell m}$ or $\rho_{\mathrm{p}}$ can be reproduced under the assumption that the \emph{true} gravitational-wave signal contains only the dominant contribution $h_{0}$ and noise is solely responsible for reproducing $h_{1}$.

We do not calculate $\rho_{1}^{\mathrm{MF}}$ directly from the data, but rather infer it from the parameter estimation results.  In particular, we determine the value of $\rho_{1}^{\mathrm{MF}}$ which gives an expected posterior distribution $p(\rho_{1} | d)$ from Eqs.~(\ref{eq:marg_like}) and (\ref{eq:analytic_posterior}) that matches the $\rho_{\ell m}$ or $\rho_{\mathrm{p}}$ distribution inferred from the parameter estimation samples. We then calculate the probability of drawing $\rho_{1}^{\mathrm{MF}}$ or larger from the expected noise distribution in the absence of higher multipoles and precession, i.e. from a chi distribution with 2 degrees of freedom. In practice, we obtain the value of $\rho_{1}^{\mathrm{MF}}$ by minimising the Jensen-Shannon divergence (JSD)~\cite{61115} between $p(\rho_{1} | d)$ and the inferred posterior. 

\section{Results}
\label{sec:results}

\begin{table}
    \begin{ruledtabular}
        \begin{tabular}{| l | c c c c |}
            Event & $\rho_{33}$ & $P_{33} (\%)$ & $\rho_{p}$ & $P_{\mathrm{p}} (\%)$ \\
            \hline
            GW190403\_051519\,\, & $1.4^{+1.3}_{-1.1}$ & - & $0.3^{+0.9}_{-0.2}$ & - \\ 
            GW190408\_181802 & $0.5^{+1.1}_{-0.5}$ & 64.0 & $1.0^{+1.8}_{-0.9}$ & 21.0 \\ 
            GW190412 & $3.5^{+0.8}_{-1.2}$ & 0.24 & $3.0^{+1.6}_{-1.5}$ & 1.3 \\ 
            GW190413\_134308 & $0.7^{+1.2}_{-0.6}$ & 62.0 & $0.7^{+1.5}_{-0.6}$ & 15.0 \\ 
            GW190413\_052954 & $0.5^{+1.1}_{-0.5}$ & 11.0 & $0.6^{+1.4}_{-0.5}$ & 40.0 \\ 
            GW190421\_213856 & $0.4^{+0.9}_{-0.4}$ & 95.0 & $0.7^{+1.4}_{-0.6}$ & 24.0 \\ 
            GW190426\_190642 & $0.8^{+1.6}_{-0.7}$ & - & $0.2^{+0.5}_{-0.2}$ & - \\ 
            GW190503\_185404 & $0.8^{+1.3}_{-0.7}$ & 21.0 & $0.8^{+1.8}_{-0.7}$ & 28.0 \\ 
            GW190512\_180714 & $1.1^{+1.1}_{-0.9}$ & 19.0 & $0.8^{+1.6}_{-0.7}$ & 59.0 \\ 
            GW190513\_205428 & $1.1^{+1.4}_{-1.0}$ & 24.0 & $0.8^{+1.6}_{-0.6}$ & 71.0 \\ 
            GW190514\_065416 & $0.4^{+1.0}_{-0.4}$ & 39.0 & $0.5^{+1.2}_{-0.4}$ & 28.0 \\ 
            GW190517\_055101 & $0.7^{+1.3}_{-0.7}$ & 39.0 & $1.0^{+2.0}_{-0.8}$ & 31.0 \\ 
            GW190519\_153544 & $2.3^{+1.5}_{-1.8}$ & 1.6 & $1.0^{+1.9}_{-0.7}$ & 13.0 \\ 
            GW190521 & $1.2^{+2.4}_{-1.1}$ & - & $0.7^{+1.4}_{-0.6}$ & - \\ 
            GW190521\_074359 & $1.0^{+1.5}_{-0.9}$ & 27.0 & $1.6^{+2.5}_{-1.2}$ & 11.0 \\ 
            GW190527\_092055 & $0.6^{+1.1}_{-0.5}$ & 75.0 & $0.7^{+1.7}_{-0.6}$ & 37.0 \\ 
            GW190602\_175927 & $0.8^{+1.4}_{-0.8}$ & 59.0 & $0.5^{+1.0}_{-0.4}$ & 100.0 \\ 
            GW190620\_030421 & $1.1^{+1.5}_{-1.0}$ & 28.0 & $0.8^{+1.7}_{-0.6}$ & 45.0 \\ 
            GW190630\_185205 & $1.0^{+1.2}_{-0.9}$ & 25.0 & $1.0^{+1.8}_{-0.8}$ & 35.0 \\ 
            GW190701\_203306 & $0.5^{+1.0}_{-0.4}$ & 83.0 & $0.5^{+1.0}_{-0.4}$ & 81.0 \\ 
            GW190706\_222641 & $1.5^{+1.5}_{-1.3}$ & 11.0 & $0.5^{+1.1}_{-0.4}$ & 51.0 \\ 
            GW190707\_093326 & $0.4^{+0.8}_{-0.3}$ & 48.0 & $0.7^{+1.5}_{-0.6}$ & 81.0 \\ 
            GW190708\_232457 & $0.4^{+0.9}_{-0.3}$ & 77.0 & $0.7^{+1.5}_{-0.6}$ & 58.0 \\ 
            GW190719\_215514 & $0.7^{+1.2}_{-0.6}$ & 69.0 & $0.6^{+1.5}_{-0.5}$ & 52.0 \\ 
            GW190720\_000836 & $0.5^{+0.9}_{-0.4}$ & 46.0 & $0.6^{+1.2}_{-0.5}$ & 49.0 \\ 
            GW190725\_174728 & $0.5^{+1.0}_{-0.5}$ & - & $1.0^{+1.9}_{-0.8}$ & - \\ 
            GW190727\_060333 & $0.5^{+1.2}_{-0.4}$ & 76.0 & $0.7^{+1.6}_{-0.6}$ & 36.0 \\ 
            GW190728\_064510 & $0.5^{+1.2}_{-0.4}$ & 34.0 & $0.7^{+1.3}_{-0.6}$ & 40.0 \\ 
            GW190731\_140936 & $0.5^{+1.1}_{-0.4}$ & 49.0 & $0.5^{+1.3}_{-0.4}$ & 64.0 \\ 
            GW190803\_022701 & $0.4^{+0.9}_{-0.4}$ & 77.0 & $0.6^{+1.4}_{-0.5}$ & 43.0 \\ 
            GW190805\_211137 & $0.6^{+1.1}_{-0.5}$ & - & $0.5^{+1.2}_{-0.4}$ & - \\ 
            GW190814 & $6.2^{+1.3}_{-1.5}$ & $1.7\times 10^{-6}$ & $1.8^{+1.6}_{-1.2}$ & 21.0 \\ 
            GW190828\_063405 & $0.5^{+1.0}_{-0.4}$ & 51.0 & $0.9^{+1.6}_{-0.8}$ & 39.0 \\ 
            GW190828\_065509 & $1.2^{+1.0}_{-1.0}$ & 8.5 & $1.0^{+1.9}_{-0.8}$ & 24.0 \\ 
            GW190910\_112807 & $0.6^{+1.3}_{-0.6}$ & 15.0 & $0.8^{+1.6}_{-0.7}$ & 26.0 \\ 
            GW190915\_235702 & $0.5^{+1.0}_{-0.5}$ & 51.0 & $1.5^{+2.4}_{-1.2}$ & 3.7 \\ 
            GW190916\_200658 & $0.8^{+1.3}_{-0.7}$ & - & $0.5^{+1.2}_{-0.4}$ & - \\ 
            GW190924\_021846 & $0.5^{+1.0}_{-0.5}$ & 18.0 & $0.5^{+1.1}_{-0.5}$ & 100.0 \\ 
            GW190925\_232845 & $0.3^{+0.8}_{-0.3}$ & - & $0.7^{+1.3}_{-0.6}$ & - \\ 
            GW190926\_050336 & $0.8^{+1.4}_{-0.7}$ & - & $0.7^{+1.7}_{-0.6}$ & - \\ 
            GW190929\_012149 & $2.0^{+1.6}_{-1.5}$ & 4.2 & $0.9^{+2.0}_{-0.7}$ & 13.0 \\ 
            GW190930\_133541 & $0.4^{+1.0}_{-0.3}$ & 35.0 & $0.6^{+1.2}_{-0.5}$ & 44.0 \\ 
            \end{tabular}
        \end{ruledtabular}
    \caption{Table showing the SNR in the ($\ell, m$) = (3, 3) multipole moment $\rho_{33}$ and the SNR from precession $\rho_{\mathrm{p}}$ for all BBH candidates observed in GWTC-2.1~\cite{LIGOScientific:2021usb}. For each event we show two p-values; $P_{33}$ and $P_{\mathrm{p}}$ show the probability that the inferred posterior is caused by noise. Events with a smaller p-value show greater evidence for higher order multipoles and/or precession. For events where the p-value and $\rho_{33}$/$\rho_{\mathrm{p}}$ could not be calculated we add a hyphen. Where applicable we report the median values along with the 90\% symmetric credible intervals.}
    \label{tab:rho_p}
\end{table}

\begin{figure*}[t!]
        \centering
        \includegraphics[width=0.48\textwidth]{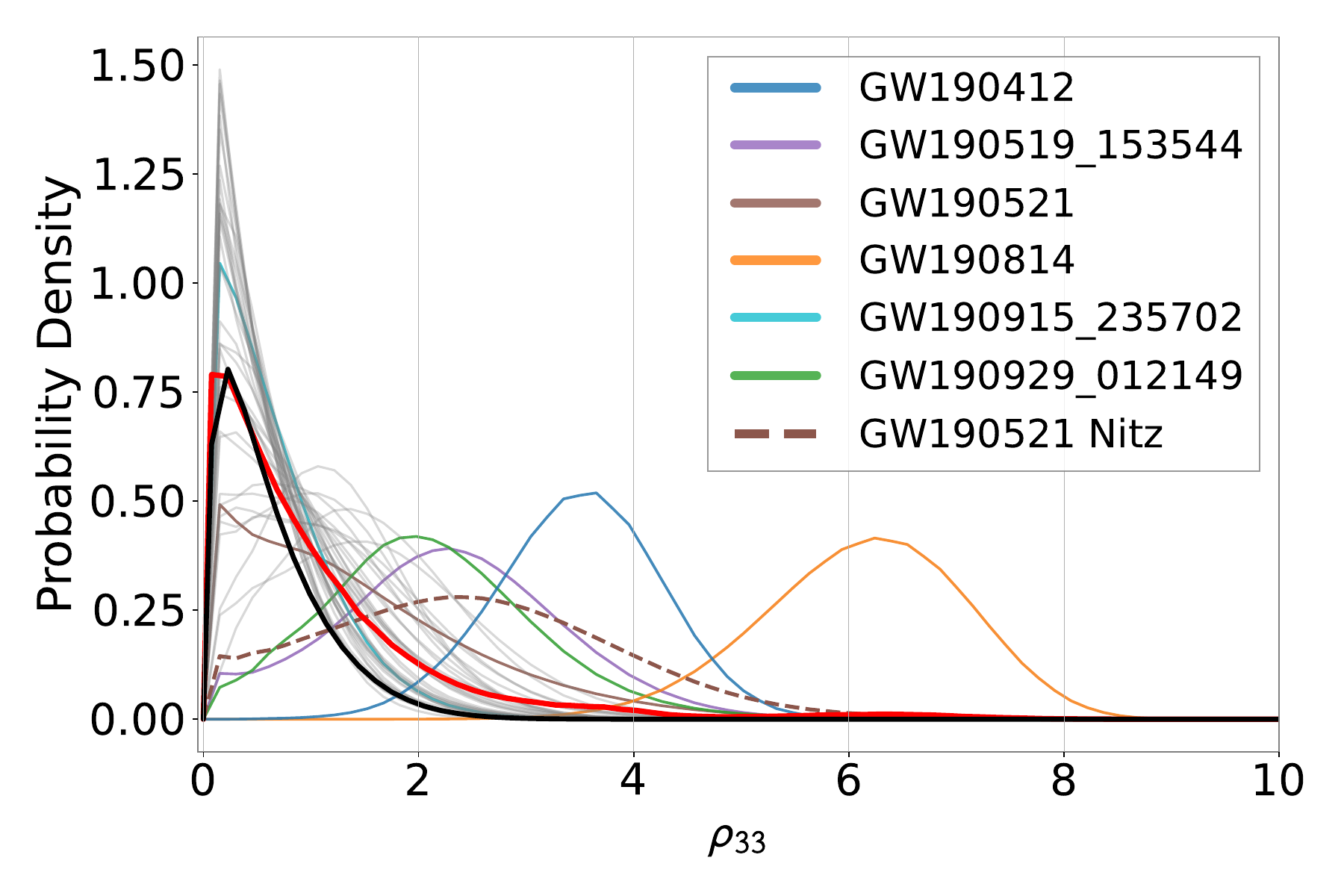}
        \includegraphics[width=0.46\textwidth]{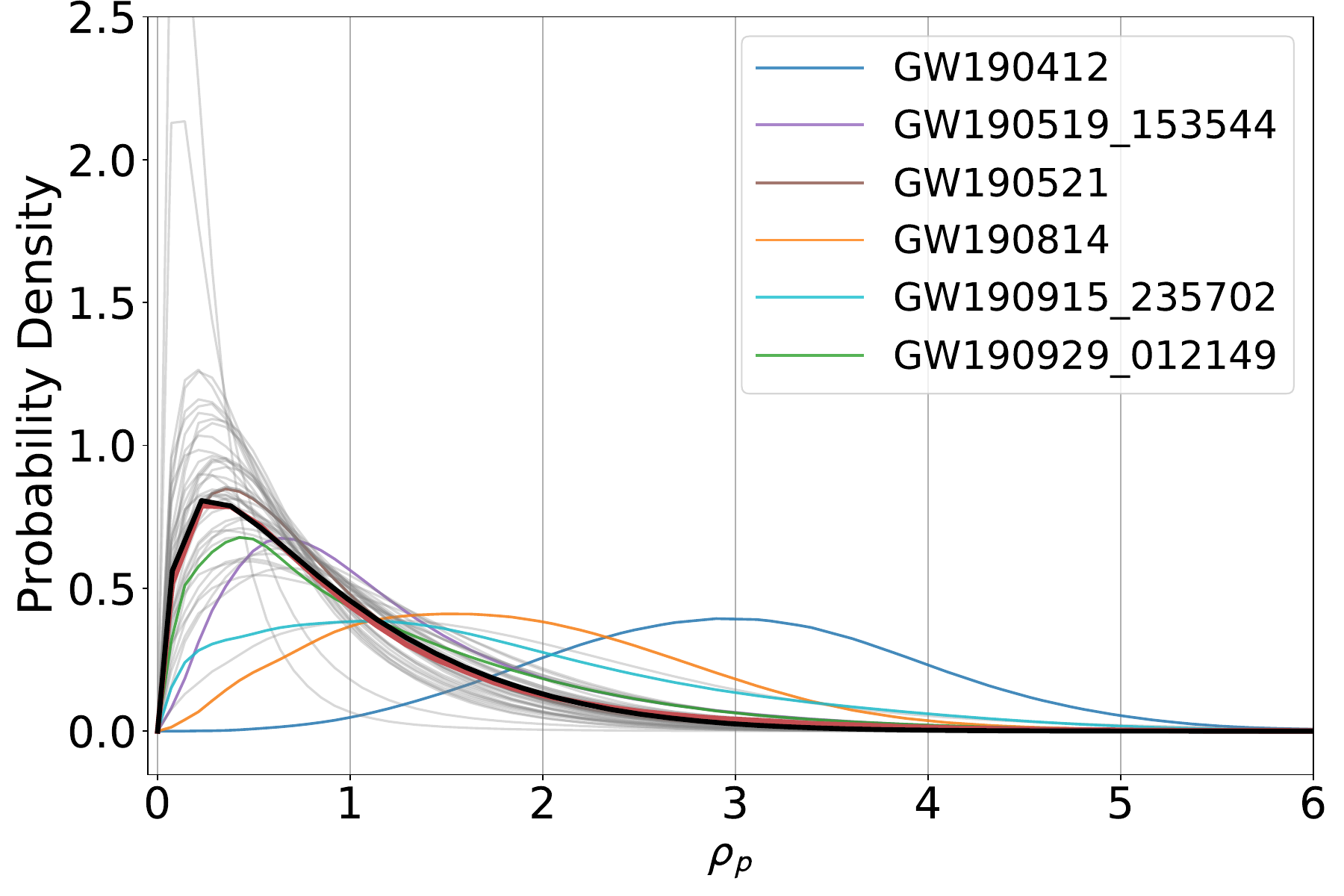}
        \caption{Plot showing the \emph{Left}: $\rho_{33}$ and \emph{Right}: $\rho_{\mathrm{p}}$ distributions for all 
        observations in the second GW catalogue (grey). In red we show the $\rho_{33}$ and $\rho_{\mathrm{p}}$ distribution averaged across events. In black we show the average of the median expected noise distribution for \emph{Left:} higher multipoles and \emph{Right:} precession. Events which are discussed in the text are colored.}
        \label{fig:rho_p}
\end{figure*}

From the publicly released posterior samples, we are able to quantify the evidence of subdominant multipole moments and precession in the observed GW signal. Our analysis finds that several candidates in GWTC-2.1 show strong evidence for the presence of subdominant multipole moments and others show marginal evidence for precession. A summary of the main results is given in Table~\ref{tab:rho_p}.

We report the observed SNR in the (3, 3) multipole and precession for all events in GWTC-2.1. Where possible, we also provide the probability that the inferred SNR in the (3,3) multipole and from precession is caused by noise, $P_{33}$ and $P_{\mathrm{p}}$ respectively. Although we calculate the SNR in the (2,1), (3,3) and (4,4) subdominant multipoles, we only report the evidence for the (3,3). This is because our analysis finds that the (3,3) multipole is the most significant sub-dominant multipole for every event except one in O3a. This is expected since, across the majority of the parameter space, the expected SNR in the (3,3) multipole is largest.  It is only binaries with a) large total mass -- where the in-band power in the (4,4) is larger -- and b) close to equal mass components -- where the (3,3) multipole vanishes~\cite{Mills:2020thr} --- that the expected SNR in the (4,4) multipole is larger. GW190910\_112807 is the sole exception, having inferred $\rho_{33} = 0.6^{+1.3}_{-0.6}$ and $\rho_{44} = 1.0^{+0.5}_{-1.0}$, both of which are consistent with noise with p-values $>10\%$. As expected, GW190910\_112807's source has close to equal mass components, $q = m_1/m_2=1.22_{-0.20}^{+0.48}$, and has relatively large total mass $M = m_{1} + m_{2} = 79.6^{+9.3}_{-9.1} M_{\odot}$.  It also has significant support for an edge-on orientation, where the relative amplitude of the (4,4) multipole is largest. 

Several events in GWTC-2.1 have an SNR in the (3,3) multipole which is clearly above the expectation for noise alone.  Indeed, the observed distribution for the population, shown in Figure~\ref{fig:rho_p}, shows a clear high-SNR tail that indicates an observation of the (3,3) multipole.  Higher multipoles have previously been identified in both GW190412 and GW190814, with their observability and their impact on parameter estimates discussed at length in previous works~\cite{LIGOScientific:2020stg, Colleoni:2020tgc, Islam:2020reh, Capano:2020dix, Abbott:2020khf}. Unsurprisingly, we see that among all events in O3a, GW190412 and GW190814 have the largest SNRs in the (3, 3) multipole, with $\rho_{33} = 3.5^{+0.8}_{-1.2}$ and $\rho_{33} = 6.2^{+1.3}_{-1.5}$ respectively. For both events, the observed $\rho_{33}$ is unlikely to be caused by noise since there is an approximately 1 in 400 and 1 in $6\times10^{7}$ chance that the observed distribution is caused by noise.

The events GW190519\_153544 and GW190929\_012149 show marginal evidence for an observable signal in the (3, 3) multipole.  The observed SNRS are $\rho_{33} = 2.3^{+1.5}_{-1.8}$ and $\rho_{33} = 2.0^{+1.6}_{-1.5}$ respectively with associated p-values of $P_{33} = 1.6\%$ and $P_{33} = 4.2\%$ respectively.  While a 1.6\% p-value would be significant in a single trial, this is consistent with expectations when considering over thirty events from O3a.

The evidence for precession is weaker than for higher multipoles.  As shown in Figure \ref{fig:rho_p}, there are only two events which show any significant deviation from the average expected noise distribution.

GW190412 shows the strongest evidence for precession with $\rho_{\mathrm{p}} = 3.0^{+1.6}_{-1.5}$ and p-value 1.3\%. Although significant in a single trial, this is consistent with expectations in a population of thirty events. While GW190814 has the second-largest precession SNR, $\rho_{\mathrm{p}} = 1.8^{+1.6}_{-1.2}$, the observed distribution is consistent with zero precession: $P_{\mathrm{p}} = 21\%$. We also find that GW190915\_235702 shows marginal evidence for measurable precession with $\rho_{p} = 1.5^{+2.4}_{-1.2}$ and $P_{\mathrm{p}} = 3.7\%$.

For the other events in GWTC-2.1, there is no evidence for precession since the inferred $\rho_{\mathrm{p}}$ is consistent with noise: $P_{\mathrm{p}} > 10\%$.%
\footnote{We note that two events, GW190602\_175927 and GW190924\_175927, have precession p-values $P_{\mathrm{p}} = 100\%$.  The reason that these events have such high p-values is because the inferred $\rho_{\mathrm{p}}$ is significantly lower than expected.  This arises because $\rho_p$ is calculated from an analysis incorporating both precession and higher harmonics, while the prior is generated from an analysis lacking both precession and higher harmonics.
In both cases, the inclusion of higher harmonics significantly improves the estimate of the mass ratio and means that the prior distribution (of both mass ratio and $\rho_p$) is slightly different than the results of the precessing, higher-harmonic analysis.} %
The lack of observable precession does not necessarily mean that most events in GWTC-2.1 have aligned-spins but rather that if the binaries were precessing, the imprint of precession on the observed signal is not strong enough to be observed with the current detector sensitivities.

Since GW190412, GW190814, GW190915\_235702, GW190519\_153544 and GW190929\_01214 all show at least some evidence for the (3, 3) multipole or precession, we discuss these events in more detail in Section~\ref{sec:individual_gws}. We also discuss GW190521 since is has the largest inferred in-plane spin of all events in GWTC-2.1.

\subsection{The population as a whole}

\begin{figure}[t!]
        \centering
        \includegraphics[width=0.48\textwidth]{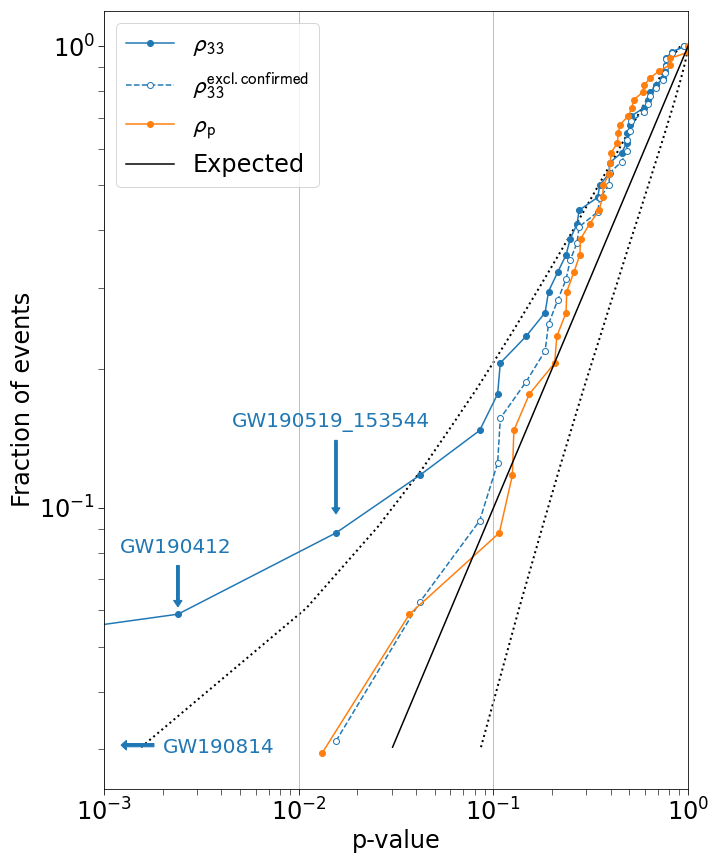}
        \caption{The cumulative distribution of p-values for the evidence of subdominant multipole moments (solid blue) and precession (orange) for the gravitational-wave candidates observed in GWTC-2.1. The solid black line indicates the expected distribution of p-values under a no-evidence hypothesis with the corresponding 90\% uncertainty band shown by the dotted black lines. The blue dashed line shows the cumulative distribution of p-values for the evidence of subdominant multipole moments when GW190412 and GW190814 are excluded from the population. The p-values for the evidence of subdominant multipole moments for GW190412, GW190519\_153544 and GW190814 are labelled.
        }
        \label{fig:pvalues}
\end{figure}

First, we discuss the evidence for the (3, 3) multipole and precession in the population of observed GW candidates. In Figure~\ref{fig:pvalues} we plot the cumulative distribution of p-values for the (3, 3) multipole and for precession. If there was no-evidence for the (3, 3) multipole and/or precession in the population, we expect to observe a uniform distribution of p-values.

Our analysis finds strong evidence for the presence of the (3, 3) multipole in the population of GW candidates since the cumulative distribution of p-values lies outside of the 90\% confidence interval of the no-signal hypothesis. We find that, as expected, both GW190412 and GW190814 are influential in this observation.  When these two events are removed from the population, the cumulative distribution of p-values lies within the 90\% confidence interval.

For the case of precession, we find no significant evidence for precession in the population of GW candidates and, overall, the population distribution for $\rho_{p}$ is consistent with that expected from a non-precessing population.  We note that the existence of misaligned spins has been inferred for this population, see e.g. Ref.~\cite{abbott2020gwtc2prop}.  However, these results are not necessarily incompatible as the requirement for misaligned spins can be driven by binaries with negative aligned spin, rather than in-plane spins.

For both the (3,3) multipole and precession analyses, we observe an excess of low significance events which have lower p-values than expected, see the upper-right of Figure \ref{fig:pvalues}.  Indeed, at around a p-value of 0.5, the excess lies outside the 90\% region for both $\rho_{33}$ and $\rho_{p}$. Although apparent in both analyses, the excess is more significant for the precession analysis. We have not definitively identified the cause of this excess, but note  several possible explanations.  First, as discussed in Appendix \ref{ap:samples}, for the majority of events the informed prior for the precession analysis is obtained using a waveform model that lacks both precession and higher multipoles.  Ideally, we would use results an analysis which differed only in its treatment of precession, but none is available in the public data.  This could lead to an incorrect estimate of the informed prior and consequently the p-value.  Second, as discussed in Section \ref{sec:noise_rho}, when deriving the form of the expected posterior in Eq.~\ref{eq:analytic_posterior}, we assume that the masses and aligned spins are measured exactly from the (2,2) waveform.  In reality, this is not the case and could lead to small differences in the inferred p-values.  Finally, the result could be genuine, in that this is a genuine statistical fluctuation -- it is not unreasonable to observe an excess outside of the 90\% confidence interval. In fact, the generic expectation is that both the (3,3) and precessing harmonics will usually be present in the data but buried in the noise. The combination of a sub-threshold signal and noise is more likely to lead to excesses such as the one we observe here. It is worth noting, however, that this excess at high p-values does not impact our overall conclusions or the robustness of results at low p-values.  

\subsection{Individual GW candidates} \label{sec:individual_gws}

Here, we discuss in more detail all GW candidates for which the p-value for higher multipoles or precession (or both) is less than 5\%.

\subsubsection{GW190814}
\label{sec:0814}

\begin{figure*}[t!]
        \centering
        \includegraphics[width=0.98\textwidth]{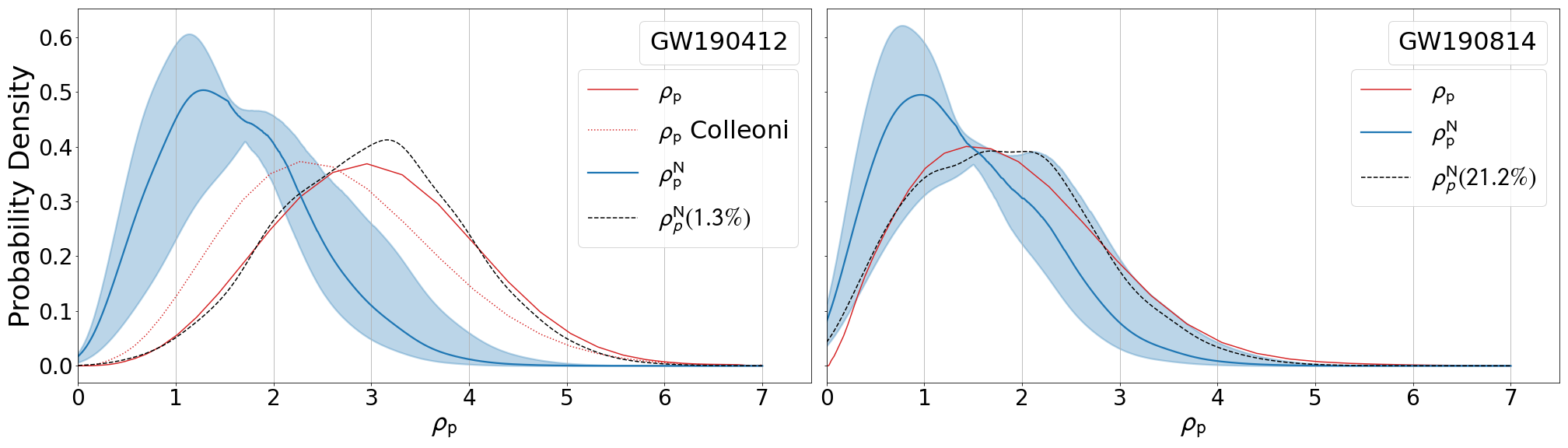}
        \caption{$\rho_{\mathrm{p}}$ distributions for \emph{Left}: GW190412 and \emph{Right}: GW190814. The blue line shows the expected distribution of $\rho_{\mathrm{p}}$ in a stretch of noisy data under the assumption that the source is non-precessing, $\rho_{\mathrm{p}}^{\mathrm{N}}$. The blue shaded region shows the $1\sigma$ uncertainty of $\rho_{\mathrm{p}}^{\mathrm{N}}$  and the black dashed line shows the expected $\rho_{p}^{\mathrm{N}}$ distribution in \emph{Left}: 1.3\% and \emph{Right}: 21.2\% of cases. The black dashed line was calculated by finding the value of $\rho_{1}^{\mathrm{MF}}$ which minimized the Jensen-Shannon divergence between the $\rho_{\mathrm{p}}^{\mathrm{N}}$ and the inferred $\rho_{\mathrm{p}}$ posterior. The dashed red line in the \emph{Left panel} show the inferred $\rho_{\mathrm{p}}$ distributions calculated using the samples from Colleoni \emph{et al.}~\cite{Colleoni:2020tgc}.}
        \label{fig:rho_p_events}
\end{figure*}

GW190814 is the most unequal mass ratio binary observed in O3a. The component masses were inferred to be $23.2^{+1.1}_{-1.0}\, M_{\odot}$ and $2.59^{+0.08}_{-0.09}\, M_{\odot}$ which makes the secondary component mass either the heaviest neutron star or lightest black hole ever recorded. Previously, GW190814 was found to have significant evidence for subdominant multipole moments owing to the unequal component masses~\cite{Abbott:2020khf}.

Unsurprisingly, we infer that GW190814 has the most significant measurement of $\rho_{33}$ in O3a. We find that the inferred $\rho_{33}$ measurement is inconsistent with noise since the associated p-value is significantly smaller than $1\%$ and is the smallest for any event in O3a, $P_{33} = 1.7\times10^{-6}\%$.

Although GW190814 has the second largest $\rho_{\mathrm{p}}$, there is minimal for precession in the observed GW signal.  The inferred $\rho_{\mathrm{p}}$ measurement can be reproduced from noise in $21\%$ of cases, as shown in Figure~\ref{fig:rho_p_events}.
GW190814 is an example where a large inferred $\rho_{\mathrm{p}}$ does not correlate with observable precession.  Since this binary has highly unequal masses, a small in-plane spin would lead to observable precession.  Therefore, our prior belief is that large values of $\rho_{p}$ are quite possible --- this differs from an equal mass binary for which it is very unlikely to obtain a large $\rho_{p}$.  Consequently, GW190814's inferred $\rho_{\mathrm{p}}$ distribution is well contained within the $1\sigma$ noise uncertainty, see Figure~\ref{fig:rho_p_events}. GW190814 therefore demonstrates the efficacy of our algorithm for inferring the presence of precession.

The lack of observable precession in GW190814 implies that it's source is either non-precessing or we are unable to observe the precession at current detector sensitivities. This is a similar conclusion to that stated in Ref.~\cite{Pratten:2020igi} which highlighted that a precessing GW190814-like system with in-plane spin $0 < \chi_{\mathrm{p}} < 0.1$ is indistinguishable from a non-precessing system based on the difference in Bayesian evidence.

\subsubsection{GW190412}
\label{sec:0412}

\begin{table}[t!]
    \begin{ruledtabular}
        \begin{tabular}{| l | c c c c |}
            Analysis & $\rho_{33}$ & $P_{33} (\%)$ & $\rho_{p}$ & $P_{\mathrm{p}} (\%)$ \\
            \hline
            LVK & $3.5^{+0.8}_{-1.2}$ & 0.24 & $3.0^{+1.6}_{-1.5}$ & 1.3 \\ 
            Colleoni \emph{et al.} & $3.5^{+1.1}_{-1.2}$ & 0.030 & $2.5^{+1.8}_{-1.4}$ & 5.1 \\ 
            Nitz \emph{et al.} & $3.4^{+1.3}_{-1.3}$ & - & $2.3^{+1.8}_{-1.3}$ & - \\ 
            Zevin \emph{et al.} & $3.5^{+0.9}_{-1.2}$ & - & $2.9^{+1.7}_{-1.6}$ & - \\
            \end{tabular}
        \end{ruledtabular}
    \caption{Table as in Table~\ref{tab:rho_p} but showing only the inferred posteriors and p-values for GW190412. We compare analyses from the LVK~\cite{LIGOScientific:2020stg, ligo_scientific_collaboration_and_virgo_2021_gwtc2}, Colleoni \emph{et al.}~\cite{Colleoni:2020tgc, colleoni_marta_2020_4079188}, Nitz \emph{et al.}~\cite{Nitz:2021uxj} and Zevin \emph{et al.}~\cite{Zevin:2020gxf, zevin_2020_3900547}. We calculate $\rho_{33}$ and $\rho_{\mathrm{p}}$ for Zevin \emph{et al.} by using posterior samples obtained from the ``Model A'' analysis since the priors are the same as those used in Ref.~\cite{LIGOScientific:2020stg}. We equally combined the posterior samples obtained with the {\sc{SEOBNRv4PHM}}~\cite{Ossokine:2020kjp} and {\sc{IMRPhenomPv3HM}}~\cite{Khan:2019kot} waveform models as was done in Ref.~\cite{LIGOScientific:2020stg}. }
    \label{tab:different_0412}
\end{table}

GW190412 was the first detection of a BBH with conclusively unequal component masses: $30.1^{+4.6}_{-5.3}\, M_{\odot}$ and $8.3^{+1.6}_{-0.9}\, M_{\odot}$ and the first observation where subdominant multipole moments were clearly observed. GW190412 was also the first observation where an informative precession measurement was inferred, with the posterior deviating significantly from the prior~\cite{LIGOScientific:2020stg}. Several groups later re-analysed GW190412 and found similar results~\cite{Colleoni:2020tgc, Zevin:2020gxf, Nitz:2021uxj, Islam:2020reh}.

We infer that GW190412 has the second most significant measurement of $\rho_{33}$ in O3a, and that the inferred $\rho_{33}$ measurement is inconsistent with noise with $P_{33} = 0.2\%$.  This suggests that higher multipoles are present in the system, a result consistent with parameter estimation studies from Refs.~\cite{LIGOScientific:2020stg,Colleoni:2020tgc,Islam:2020reh}. In Table~\ref{tab:different_0412}, we present results for the significance of the (3, 3) multipole from several different analyses of GW19041, and consistently show that it is found with a significant SNR.%
\footnote{Islam \emph{et al.}~\cite{Islam:2020reh} also re-analysed GW190412 using the {\sc{NRSur7dq4}} waveform model~\cite{Varma:2019csw} but their samples are not publicly available and therefore not included in this work.}.

\begin{figure}[t!]
    \centering
    \includegraphics[width=0.48\textwidth]{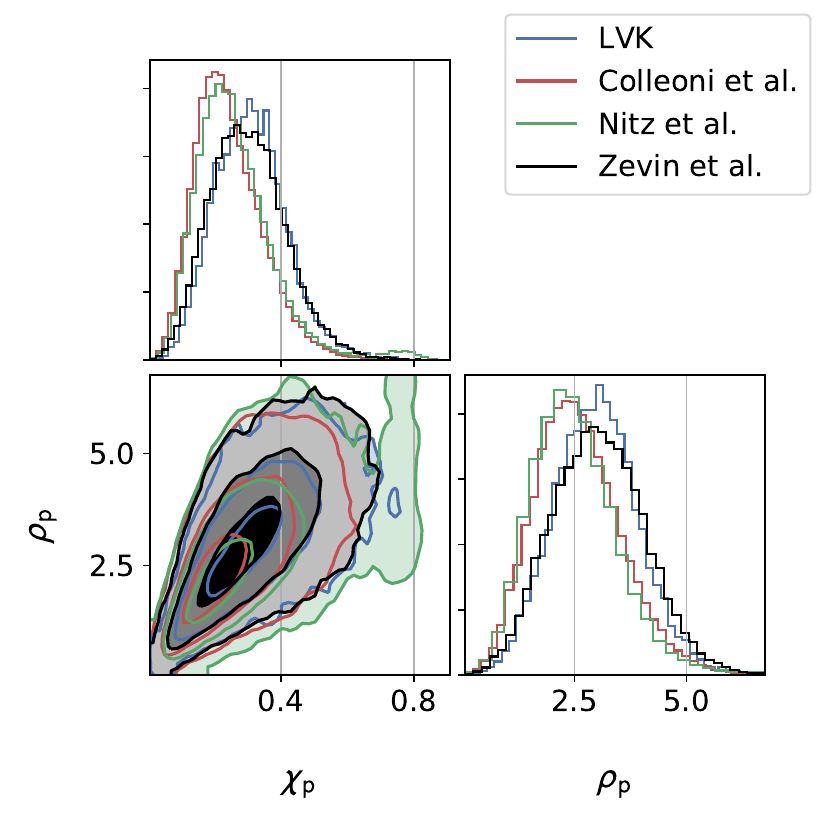}
    \caption{Corner plot comparing the inferred $\chi_{\mathrm{p}}$ and $\rho_{\mathrm{p}}$ for GW190412 from the LVK~\cite{LIGOScientific:2020stg, ligo_scientific_collaboration_and_virgo_2021_gwtc2}, Colleoni \emph{et al.}~\cite{Colleoni:2020tgc, colleoni_marta_2020_4079188}, Nitz \emph{et al.}~\cite{Nitz:2021uxj} and Zevin \emph{et al.}~\cite{Zevin:2020gxf, zevin_2020_3900547}. Shading shows the $1\sigma$, $3\sigma$ and $5\sigma$ confidence intervals.}
    \label{fig:0412_corner}
\end{figure}

We find that the evidence for precession in GW190412 is dependent on which Bayesian analysis is considered. We calculate that GW190412 shows marginal evidence for precession when using data from the initial analysis conducted by the LVK~\cite{LIGOScientific:2020stg}. We find that the inferred precession SNR can be reproduced from noise in only 1 in every 100 cases, see Figure~\ref{fig:rho_p_events}.
However, when using data produced from a re-analysis of GW190412 using the latest suite of Phenomenological waveform models ({\sc{PhenomX}}~\cite{Pratten:2020fqn, Garcia-Quiros:2020qpx, Pratten:2020ceb, Estelles:2020osj, Estelles:2021gvs})~\cite{Colleoni:2020tgc, colleoni_marta_2020_4079188}, hereafter Colleoni \emph{et al.}, GW190412 shows low evidence for precession since the inferred $\rho_{\mathrm{p}}$ is smaller than that reported by the LVK and can be reproduced from noise in 1 in every 20 cases. The smaller $\rho_{\mathrm{p}}$ is a consequence of inferring a lower in-plane spin, as shown in Figure~\ref{fig:0412_corner}.

Nitz \emph{et al.}~\cite{Nitz:2021uxj} and Zevin \emph{et al.}~\cite{Zevin:2020gxf} also performed independent analyses of GW190412, and the results for $\rho_{p}$ are included in Table~\ref{tab:different_0412} and Figure~\ref{fig:0412_corner}.  However, since they did not release results for aligned-spin waveform models,  we are unable to calculate p-values for the evidence of precession in the observed GW signal.
Nonetheless, the inferred $\rho_{\mathrm{p}}$ and $\chi_{\mathrm{p}}$ distributions from Nitz \emph{et al.} and Zevin \emph{et al.} are comparable to the measurements reported in Colleoni \emph{et al.} and the LVK respectively.

 Colleoni \emph{et al.} and Nitz \emph{et al.} both used the {\sc{IMRPhenomXPHM}} waveform model~\cite{Pratten:2020ceb} for the Bayesian inference while the LVK and Zevin \emph{et al.} used a combination of the {\sc{IMRPhenomPv3HM}}~\cite{Khan:2019kot} and {\sc{SEOBNRv4PHM}}~\cite{Ossokine:2020kjp} waveform models. This suggests that the differences we see between interpretations could either be a consequence of waveform systematics, difficulties in sampling the complex parameter space or sampler differences.

\subsubsection{GW190519\_153544 and GW190929\_012149}
\label{sec:0519_0929}

\begin{figure*}[t!]
    \centering
    \includegraphics[width=0.98\textwidth]{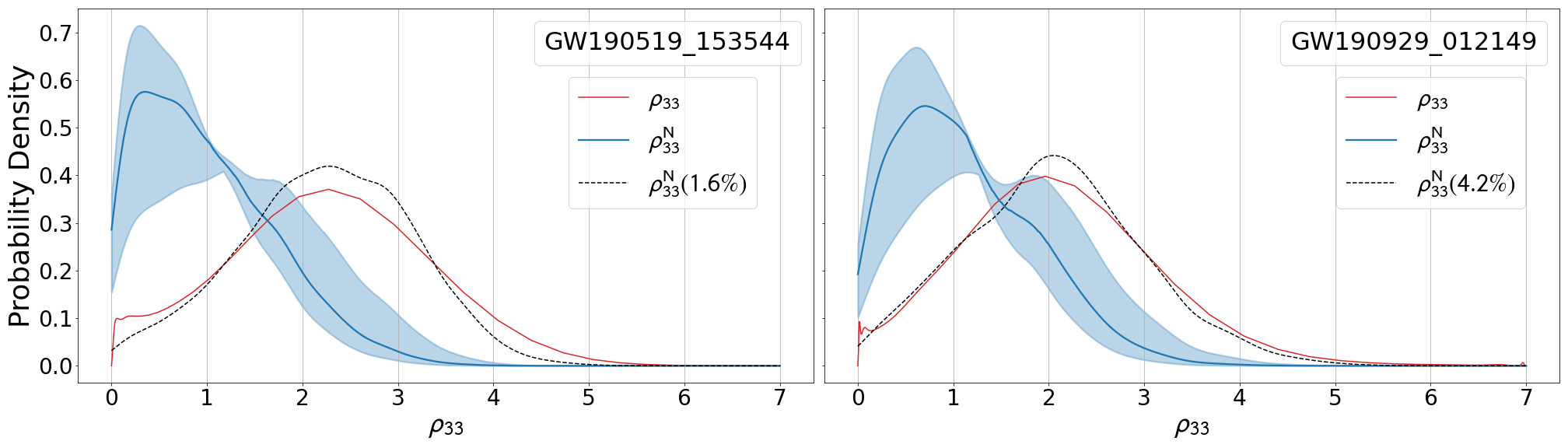}
    \caption{$\rho_{33}$ for \emph{Left}:  GW190519\_153544 and \emph{Right}: GW190929\_012149. The blue line shows the expected distribution of $\rho_{33}$ in a stretch of noisy data, $\rho_{33}^{\mathrm{N}}$. The blue shaded region shows the $1\sigma$ uncertainty of $\rho_{33}^{\mathrm{N}}$, the grey dotted line shows the expected $\rho_{33}^{\mathrm{N}}$ distribution in 10\% of cases and the black dashed line shows the expected $\rho_{33}^{\mathrm{N}}$ distribution in \emph{Left}: 1.5\% and \emph{Right}: 10\% of cases. The black dashed line was calculated by finding the $\rho_{1}^{\mathrm{MF}}$ which minimized the Jensen-Shannon divergence between the $\rho_{33}^{\mathrm{N}}$ and the inferred $\rho_{33}$ posterior.}
    \label{fig:rho_33_3events}
\end{figure*}

GW190519\_153544 originated from a binary with relatively high total mass, 50\% posterior probability for $M > 100 M_{\odot}$, and with spins preferentially aligned with the orbital angular momentum, $\chi_{\mathrm{eff}} = 0.31^{+0.20}_{-0.22}$.  GW190519\_153544 shows marginal evidence for higher order multipoles since there is a $2\%$ probability of recovering the inferred $\rho_{33}$ from noise, see Figure~\ref{fig:rho_33_3events}.  The inferred properties of the signal change significantly when higher harmonics are considered (as shown in Figure 13 of Ref.~\cite{Abbott:2020niy}).  Specifically, the mass ratio is constrained more tightly and the binary is inferred to be edge-on, rather than face-on.  Both of these effects are consistent with the presence of higher harmonics in the signal.  However, as shown in Figure~\ref{fig:pvalues}, when GW190412 and GW190814 are removed the population, the observed $\rho_{33}$ is consistent with the population expectations from a no-signal hypothesis.  

GW190929\_012149 also shows marginal evidence for higher-order multipoles, with $\rho_{33} = 2.0^{+1.6}_{-1.5}$ which is expected from noise only in 4\% of cases. Inclusion of higher harmonics in parameter recovery does lead to improved inference of the mass ratio, but has little impact on the measured orientation.  Again, while a 4\% chance might be significant for a single event, this observation is consistent with being an outlier in a population of thirty events.

Neither GW190519\_153544 or GW190929\_012149 show strong evidence for precession, as the inferred $\rho_{\mathrm{p}}$ is consistent with noise ($P_{p} \sim 13\%$ for both events).

\subsubsection{GW190915\_235702}
\label{sec:0915}

In our analysis, GW190915\_235702 has the second-largest evidence of precession in O3a with $P_{p} = 4\%$. GW190915\_235702 originated from a BBH with component masses $35.3^{+9.5}_{-6.4}\, M_{\odot}$ and $24.4^{+5.6}_{-6.1}\, M_{\odot}$.  The LVK analysis shows signs of precession in GW190915\_235702, with $\chi_{p} \approx 0.6^{+0.3}_{-0.4}$ measured to be larger than prior expectations, see Figure 11 of Ref.~\cite{abbott2020gwtc}.  In addition, when incorporating precession, the binary's inclination is constrained to be away from face-on.

GW190915\_235702 has no evidence for subdominant multipole moments with $P_{33} = 50\%$. The lack of evidence for subdominant multipole moments is consistent with the findings from the LVK analysis which found that the effect of higher modes is either negligible or subdominant to the systematics between precessing non-higher order multipole waveforms~\cite{abbott2020gwtc}.

\subsection{An investigation of GW190521}
\label{sec:0521}

GW190521 is the most massive binary contained in GWTC-2.1. An initial analysis conducted by the LVK argued that GW190521 provided the first evidence of a new population of black holes that resist straightforward interpretation as supernova remnants, with at least one black hole lying firmly in the pulsational pair-instability mass gap ($\sim 65-120 M_\odot$)~\cite{Abbott:2020tfl, Abbott:2020mjq}. It was found that GW190521 was consistent with component masses $85^{+21}_{-14}\, M_{\odot}$ and $66^{+17}_{-18}\, M_{\odot}$. Nitz \emph{et al.}~\cite{Nitz:2020mga} later challenged this view, showing that it is possible to obtain parameter estimates consistent with component masses that instead straddle this gap. Using a uniform in mass-ratio prior, GW190521's mass posterior was multi-model with additional modes at larger mass ratio, $q\sim 6$ and $q\sim 10$, and less support for equal mass ratio systems.  Constraints on the mass ratio imposed by the initial analysis ~\cite{Abbott:2020tfl, Abbott:2020mjq, abbott2020gwtc} ruled out any possibility of sampling this high mass ratio region of the parameter space. It was later discovered that the waveform approximant used by Nitz \emph{et al.} did not accurately account for possibility of transitional precession~\cite{Apostolatos:1994mx, Nitz:2021uxj}. Nitz \emph{et al.}'s alternative interpretation of GW190521 was therefore later revised in Ref.~\cite{Nitz:2021uxj} with the high mass ratio $q\sim 10$ peak no longer significantly supported, while the mode at $q\sim 6$ remained. GW190521 may therefore have originated from either a near equal mass system, where the SNR in both the (3, 3) multipole~\cite{Mills:2020thr} and precession~\cite{Green:2020ptm} are expected to be small, or an unequal mass ratio system, where it is likely that higher order multipole and precession effects could be directly measured.

\begin{figure}[t!]
    \centering
    \includegraphics[width=0.5\textwidth]{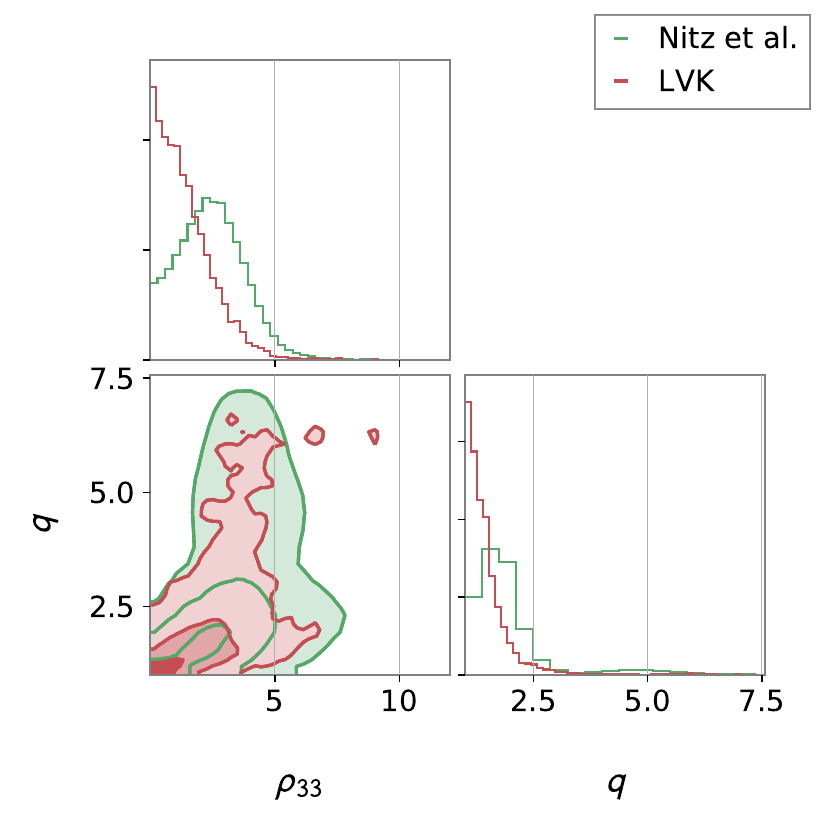}
    \caption{Corner plot showing the inferred mass ratio and $\rho_{33}$ for the reanalysis of GW190521 by Nitz \emph{et al.}~\cite{Nitz:2021uxj} compared to the results from the LVK~\cite{Abbott:2020tfl, Abbott:2020mjq, abbott2020gwtc, ligo_scientific_collaboration_and_virgo_2021_gwtc2}. Shading shows the $1\sigma$, $3\sigma$ and $5\sigma$ confidence intervals.}
    \label{fig:gw190521_corner_nitz}
\end{figure}

As might be expected, we obtain different results for the power in the (3, 3) multipole from the LVK and Nitz \emph{et al.} samples.  We infer that GW190521 has a measurable (3, 3) multipole if we use the posterior samples obtained from Nitz \emph{et al.} ($\rho_{33} = 2.4^{+2.2}_{-2.0}$) otherwise the inferred $\rho_{33}$ is consistent with Gaussian noise ($\rho_{33} = 1.2^{+2.4}_{-1.1}$). Figure~\ref{fig:gw190521_corner_nitz} shows that the inferred $\rho_{33}$ in the analyses is correlated with the mass ratio, where more unequal masses are consistent with a larger $\rho_{33}$.  Nitz \emph{et al.} infer a non-equal mass ratio system, $q=1.8^{+2.8}_{-0.6}$~\cite{Nitz:2021uxj}, while the analysis from the LVK infers $q=1.3^{+1.2}_{-0.3}$~\cite{Abbott:2020tfl, Abbott:2020mjq, abbott2020gwtc}. It is the extra likelihood from the measurement of the (3,3) multipole that is key to the Nitz~\emph{et al.} reinterpretation of GW190521 as an unequal mass binary. 

GW190521 has the largest inferred in-plane spins of any event observed in GWTC-2.1 with $\chi_{\mathrm{p}} = 0.68^{+0.26}_{-0.44}$ and $\chi_{\mathrm{p}} = 0.5^{+0.31}_{-0.33}$ as reported by the LVK~\cite{Abbott:2020tfl, Abbott:2020mjq, abbott2020gwtc} and Nitz \emph{et al.} respectively. However, GW190521 shows no evidence for precession in our analysis, with $\rho_{\mathrm{p}} = 0.7^{+1.4}_{-0.6}$ and $\rho_{\mathrm{p}} = 1.1^{+2.8}_{-0.9}$ respectively. While initially surprising, the lack of observable power in precession is a consequence of the high mass of the system.  The observed waveform contains only about four cycles (two orbits) in the detectors' sensitive frequency band.  As a result, GW190521 undergoes significantly less than one precession cycle in band and the two leading precession harmonics are highly degenerate.  Thus, there is very little power orthogonal to the dominant harmonic, leading to a small inferred value of $\rho_{\mathrm{p}}$. 

In order to explore the differences between the LVK and Nitz \emph{et al.} samples, we investigate whether the inferred $\rho_{33}$ is consistent with the observed GW strain data~\cite{vallisneri2015ligo, LIGOScientific:2019lzm} by directly extracting the SNR in the (3, 3) multipole through matched filtering~\cite{Allen:2005fk,Brown:2004vh,Babak:2012zx}.
Since the relative power in the (2, 2) and (3, 3) multipoles depends strongly on both the mass-ratio and inclination of the system~\cite[see e.g.][]{Mills:2020thr},
we can then use the extracted SNR in the (3, 3) multipole to identify the region of
parameter space consistent with the observed GW strain data and compare to the two distinct parameter estimation results.  Matched filtering is a standard procedure when searching for GW from binary mergers~\cite[see e.g.][]{Cannon:2011vi,
Privitera:2013xza,Messick:2016aqy,Hanna:2019ezx,
Sachdev:2019vvd,Usman:2015kfa,
Nitz:2017svb,Nitz:2018rgo,pycbc-software,2016CQGra..33q5012A,SPIIR2,
2012CQGra..29w5018L,2018CoPhC.231...62G}, however in most cases only the SNR in the dominant (2, 2) multipole is typically calculated~\cite[see e.g.][]{abbott2019gwtc, abbott2020gwtc, Venumadhav:2019tad, nitz20202, Nitz:2021uxj}. 

In Section \ref{sec:calc_rho}, we argued that the impact of both precession and higher harmonics will be sub-leading and therefore we can filter for each independently.  To do so, we first identify the point in the aligned-spin parameter space which gives the largest network SNR.  Then, using these parameters, we calculate the waveform for the higher multipoles, (3,3), (4,4) and (2,1), as well as the leading order precession correction.  Since the other harmonics are not orthogonal to the (2,2), we first project the waveform onto the space orthogonal to the (2,2) before filtering.    This is particularly significant for (2,1) harmonic, which has a 0.7 overlap with the (2,2).  The two leading precession harmonics are so close to degenerate (with $>95\%$ overlap between them) that we cannot reliably evaluate the precession SNR.  We then filter the orthogonal parts of each harmonic against the data from each detector and calculate the complex SNR in the harmonic at the coalescence time. To calculate the network SNR in each harmonic we project to the space where the relative amplitudes in each detector are consistent with the leading (2,2) harmonic. Although this does not take into account each detector's response function and their different PSDs, our simplification only introduces at most a 5\% error in the inferred higher multipole SNR.

\begin{figure}[t!]
    \includegraphics[width=0.5\textwidth]{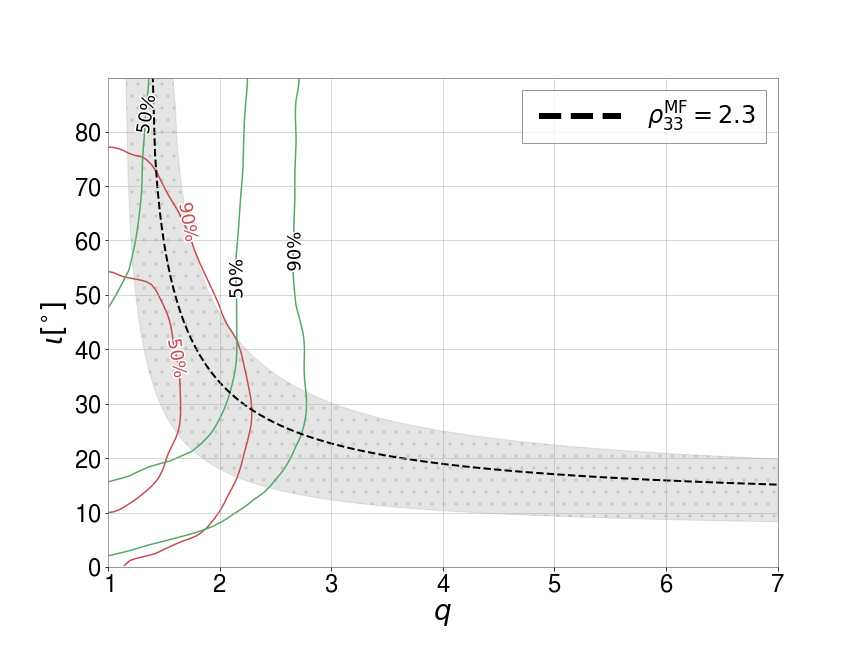}
    \caption{Region of the inclination $\iota$ and mass ratio $q$ parameter space consistent with a matched filter SNR in the (3, 3) multipole $\rho_{33}^{\mathrm{MF}} = 2.3$ (dashed black line) and a 1$\sigma$ uncertainty band (shaded grey region). Red and green contours show the 50\% and 90\% credible intervals consistent with the LVK analysis~\cite{Abbott:2020tfl, Abbott:2020mjq, abbott2020gwtc, ligo_scientific_collaboration_and_virgo_2021_gwtc2} and the Nitz \emph{et al.} analysis~\cite{Nitz:2021uxj} of GW190521 respectively.}
    \label{0521_snr_timeseries}
\end{figure}

We obtain a matched filter SNR of $\rho_{33} = 2.3$ using the above procedure.  The relative SNR in the (2,2) and (3,3) harmonics depends upon the orientation of the signal --- the SNR of the (3,3) scales with $\sin \iota$ relative to the (2,2) --- as well as the mass ratio -- with the relative SNR of the (3,3) larger for more unequal masses.  Consequently, we can identify a region of parameter space consistent with these SNRs.  The result is shown in 
Figure~\ref{0521_snr_timeseries}.  The band is broadly consistent with the Nitz \emph{et al.} results at small mass ratios, while the LVK results prefer lower mass ratios --- as might be expected due to the smaller inferred value of $\rho_{33}$.  We note that our analysis is \textit{only} using the relative SNRs in the two modes, so cannot be expected to fully reproduce the system parameters.  Higher mass ratios will be excluded by the fact that the (2,2) waveform is not a good match with the data.

We find limited power in the (4,4) harmonic, with $\rho_{44} = 1.2$, which is consistent with noise fluctuations.  However, there is significant power in the (2,1) harmonic, with $\rho_{21} = 3.4$.  This is \textit{inconsistent} with the known amplitude of the (2,1) harmonic across the parameter space: there is no combination of mass ratio and inclination that would give sufficient power in the (2,1) harmonic to yield this SNR. The observed $\rho_{21}$ is also unlikely to be a noise fluctuation since we are unable to reproduce $\rho_{21} > 3.4$ when performing the above procedure on the best matching template injected into 100 different realizations of Gaussian noise. This therefore suggests that the significant $\rho_{21}$ may be a sign of physics which isn't included in the waveform model. Several alternatives to precession have been suggested for this system, 
including possible evidence for eccentricity~\cite{Romero-Shaw:2020thy, Gayathri:2020coq} and head-on
collisions~\cite{CalderonBustillo:2020odh}.

The second, high mass ratio peak observed in Nitz \emph{et al.} is inconsistent with with 
the extracted matched filter SNR in the (3, 3) multipole.  The secondary peak correspond to binaries with $q \sim 6$, large, partially mis-aligned spins viewed close to edge-on.  For such a system, all five of the precession harmonics are important and, therefore, our aligned spin analysis with higher harmonic and precession corrections is not valid in that region of the parameter space.  

Understanding if GW190521 has a measurable (3, 3) multipole is key for understanding the binary's formation mechanism. If GW190521 has a measurable (3,3) multipole, it is unlikely to originate from an equal mass system. The preferred formation mechanism has been investigated previously with some authors suggesting that GW190521 may be a result of a hierarchical merger~\cite{Kimball:2020qyd} (although the initial LVK analysis found no conclusive evidence that GW190521 resulted from a hierarchical merger~\cite{Abbott:2020mjq}). An unequal mass ratio provides evidence that one component is the result of a previous merger~\cite{Kimball:2020qyd}. Possible evidence for eccentricity, or a head-on collision, would bolster a hierarchical interpretation ~\cite{Romero-Shaw:2020thy, Gayathri:2020coq}.

\section{Discussion} 
\label{sec:discussion}

We calculated the inferred SNR in the subdominant multipole moments, $\rho_{lm}$ (for $(\ell, |m|) \in \{(2,1), (3, 3), (4, 4)\}$, and from precession, $\rho_{\mathrm{p}}$, for all BBH candidates in GWTC-2.1 that were observed during O3a. We determined which events show evidence for subdominant multipole moments and precession by comparing the inferred SNRs with predicted distributions expected from noise alone. We found that most BBHs in O3a show minimal evidence for subdominant multipole moments, but there are a two notable exceptions. GW190412 and GW190814 show significant evidence for a (3,3) multipole, while GW190519\_153544 and GW190929\_012149 show marginal evidence for the (3, 3) multipole.  We also found that no BBH observed in O3a shows significant evidence for higher order multipole content beyond $\ell = 3$. 
We found that most BBHs in O3a show no evidence for precession. However, we found that GW190412 may have originated from a precessing binary system, with the observed result unlikely to be due to noise alone.  However, when viewed as part of the population of events from GWTC-2.1, the observation is consistent with expectations from noise fluctuations.  GW190915\_235702 shows marginal evidence for precession, which is again not significant when viewed from a population perspective. 

The interpretation of GW190521 is more complex, and is dependent upon the parameter estimation results which are used.  The LVK analysis shows no evidence for higher harmonics or precession, while the Nitz \emph{et al.} analysis shows power in the (3,3) harmonic.  For this event, we were also able to directly calculate the matched filter SNRs directly from the data and show that we do obtain power in the (3,3) harmonic, which lends strength to the argument that the binary had unequal masses.  However, we also find significant SNR in the (2,1) harmonic, which is inconsistent with physical expectations.  One explanation for this might be that GW190521 originated from an eccentric merger \cite{Romero-Shaw:2020thy, Gayathri:2020coq} in which case the waveform model we used would not contain the relevant physics.

The method we have presented is straightforward, and clearly identifies the evidence for subdominant multipole moments and precession from the observed GW signals.  As demonstrated in our analysis of GW190521, it is possible to calculate the SNR in higher-multipoles and precession directly from the data --- rather than using the results of parameter estimation analyses as we have done for the rest of the events in the paper.  This opens up the possibility of using the observed SNR in higher multipoles and precession to infer the properties of the observed binary, in advance of a detailed parameter estimation analysis.  A similar approach has been suggested proposed in Ref.~\cite{Payne:2019wmy}, where the authors demonstrated that re-weighting posteriors inferred with a (2,2) only waveform model based on the full likelihood gave results that closely match those obtained from an analysis with waveform models including higher order multipoles.  In principle, this should enable the estimation of parameters, including the effects of precession and higher multipoles,  using posteriors computed with a simpler waveform model supplemented by the measured SNRs in higher multipoles and precession.  In the future we wish to expand this method and calculate the SNR in the second, sub-dominant GW polarization as the clear observation of two polarizations is also important for breaking  degeneracies between parameters~\cite{Usman:2018imj}. 

\section{Acknowledgements}
We are grateful to Duncan Brown, Davide Gerosa, Bernard Schutz and Vivien Raymond for discussions during C. Hoy's and C. Mills's Ph.D. defences where this work was first presented. We also thank Mark Hannam, Alex Nitz and Jonathan Thompson for useful discussions and N V Krishnendu for comments on this
manuscript. This work was supported by Science and Technology Facilities Council (STFC) grant ST/N005430/1 and European Research Council (ERC)
Consolidator Grant 647839 and we are grateful for computational
resources provided by Cardiff University and LIGO Laboratory and supported by STFC grant ST/N000064/1 and National Science Foundation Grants PHY-0757058 and PHY-0823459.

This research has made use of data, software and/or web tools obtained from the Gravitational Wave Open Science Center (\href{https://www.gw-openscience.org}{https://www.gw-openscience.org}), a service of LIGO Laboratory, the LIGO Scientific Collaboration and the Virgo Collaboration. LIGO is funded by the U.S. National Science Foundation. Virgo is funded by the French Centre National de Recherche Scientifique (CNRS), the Italian Istituto Nazionale della Fisica Nucleare (INFN) and the Dutch Nikhef, with contributions by Polish and Hungarian institutes.

Plots were prepared with Matplotlib \citep{2007CSE.....9...90H}, Corner (\href{https://corner.readthedocs.io/en/latest}{https://corner.readthedocs.io})~\citep{corner} and {\sc{PESummary}} ~\citep{Hoy:2020vys}. Functions within {\sc{PyCBC}}~\cite{pycbc-software} were used to perform the matched filtering described in Section~\ref{sec:0521} and {\sc{LALSuite}}~\cite{lalsuite}, {\sc{NumPy}}~\cite{harris2020array} and {\sc{Scipy}}~\cite{2020SciPy-NMeth} were used during the analysis.

\appendix

\section{Posterior samples used}
\label{ap:samples}

For all calculations we used posterior samples re-weighed to a flat-in-comoving-volume prior to remain consistent with the results in Refs~\cite{abbott2020gwtc, LIGOScientific:2021usb}. For the majority of events we used the same posterior samples as those published in GWTC-2 and GWTC-2.1 (the ``PublicationSamples'' and ``PrecessingSpinIMRHM\_comoving'' datasets respectively). In cases where these datasets did not correspond to samples obtained with a precessing higher-order multipole approximant we used the ``C01:SEOBNRv4PHM'' dataset which includes posterior samples obtained with the {\sc{SEOBNRv4PHM}}~\cite{Ossokine:2020kjp} (precessing and higher-order multipole) approximant for both analyses\footnote{This included: GW190707\_093326, GW190720\_000836, GW190728\_064510, GW190915\_235702, GW190924\_021846, GW190929\_012149, GW190930\_133541, see Table VIII of Ref.~\cite{abbott2020gwtc}}.

Since we calculate the inferred $\rho_{\ell m}$ and $\rho_{\mathrm{p}}$ with samples obtained from a precessing higher order multipole waveform model, we calculate the \emph{informed} prior using samples obtained with a precessing non-higher order multipole and aligned-spin higher order multipole waveform model in order to ensure that the noise distribution is not biased by the absence of precession and higher order multipoles respectively. Although both Ref.~\cite{abbott2020gwtc} and Ref.~\cite{LIGOScientific:2021usb} performed parameter estimation using multiple models, Ref.~\cite{LIGOScientific:2021usb} only analysed each candidate with precessing higher order multipole waveform models while Ref.~\cite{abbott2020gwtc} analysed each candidate with aligned-spin and precessing waveform models, see Table VIII of~\cite{abbott2020gwtc}. 

Due to the lack of samples, we are unable to calculate an \emph{informed} prior, and hence noise distribution, for candidates specific to Ref.~\cite{LIGOScientific:2021usb}. For candidates described in Ref.~\cite{abbott2020gwtc} we were able to use samples obtained with a precessing non-higher order multipole waveform model (the ``PrecessingSpinIMR'' dataset) to calculate the \emph{informed} prior for $\rho_{\ell m}$ but, because not every candidate was analysed with an aligned-spin higher order multipole waveform model, we generally used samples obtained with an aligned-spin non-higher order multipole waveform model to calculate the \emph{informed} prior for $\rho_{\mathrm{p}}$ (the ``AlignedSpinIMR'' dataset). We expect that using an aligned-spin non-higher order multipole waveform model will not cause a significance difference in the obtained noise distribution for the absence of precession, since for the majority of cases the power from higher order multipoles is expected to be small and therefore parameter estimates comparable. For GW190412 and GW190814, both of which exhibit strong evidence for subdominant multipole moments~\cite{LIGOScientific:2020stg,Abbott:2020khf}, we were able to use samples obtained with an aligned-spin higher order multipole waveform model (the ``AlignedSpinIMRHM'' dataset) to calculate the expected noise distribution for the absence of precession.

\bibliographystyle{unsrt}
\bibliography{references}
\end{document}